\documentclass[10pt,letterpaper]{article} 

\usepackage{opex3}

\newcommand{\beq}{\begin{equation}}
\newcommand{\eeq}{\end{equation}}
\newcommand{\bdm}{\begin{displaymath}}
\newcommand{\edm}{\end{displaymath}}

\begin{document}

\title{Large-angle scattered light measurements for quantum-noise filter cavity design studies}

\author{Fabian Maga\~{n}a-Sandoval$^{1*}$, Rana Adhikari$^2$, Valera Frolov$^3$, Jan Harms$^2$, Jacqueline Lee$^1$, Shannon Sankar$^{4,5}$, Peter R. Saulson$^6$ and Joshua R.~Smith$^1$}

\address{
$^1$Department of Physics, California State University Fullerton, Fullerton, CA 92831, USA \\
$^2$LIGO Laboratory - California Institute of Technology, Pasadena, CA 91125, USA \\
$^3$LIGO - Livingston Observatory, Livingston, LA 70754, USA \\
$^4$LIGO Laboratory - Massachusetts Institute of Technology, Cambridge, MA 02139, USA \\
$^5$Department of Physics, University of Colorado, Boulder, CO 80309-0440, USA \\
$^6$Department of Physics, Syracuse University, Syracuse, NY 13244, USA \\
}

\email{$^*$ fabianmagana@csu.fullerton.edu}


\begin{abstract}
Optical loss from scattered light could limit the performance of quantum-noise filter cavities being considered for an upgrade to the Advanced LIGO gravitational-wave detectors. This paper describes imaging scatterometer measurements of the large-angle scattered light from two high-quality sample optics, a high reflector and a beam splitter. These optics are each superpolished fused silica substrates with silica:tantala dielectric coatings. They represent the current state-of-the art optical technology for use in filter cavities. We present angle-resolved scatter values and integrate these to estimate the total scatter over the measured angles. We find that the total integrated light scattered into larger angles can be as small as 4\,ppm.  
\end{abstract}


\bibliographystyle{osajnl}
\bibliography{references}



\section{Introduction} 

A second generation of gravitational-wave (GW) detectors is scheduled to start operation within the next 5 to 10 years \cite{LSC2009a,aVir2009,KuEA2010}. Quantum noise of the light is expected to limit the sensitivity of these detectors over a wide range of frequencies. An R\&D plan has been defined that addresses the possibility of upgrading the Advanced LIGO detector after some years of operation \cite{LSC2011}. The upgrades will further increase the significance of quantum noise at frequencies above 10\,Hz. Advanced techniques to mitigate quantum noise like the implementation of squeezed light are currently being tested in large-scale interferometers~\cite{LSC2011a}. 

To optimize the benefit from squeezed light and also to minimize the effect of optical radiation-pressure noise, it will be necessary to optically filter squeezed-light fields \cite{KLMTV2001,HaEA2003} (e.~g.~with triangular cavities). The foremost challenge when using squeezed light is to reduce the optical losses inside the interferometer. The benefit from filter cavities depends strongly on the cavity round-trip loss that can be realized. Filter-loss requirements depend on the cavity length and the main interferometer configuration, but for a 50\,m long filter cavity, the round-trip loss should not exceed a value of about 20\,ppm.
\begin{table}[ht!]
\begin{center}
\begin{tabular}{|l|l|l|l|}
\hline
Length [m] & Beam radii [mm] & Lpm [ppm]  & Year \\
\hline\hline
10 & 1.9/2.1 & 60 & 84 \cite{AFM1984} \\
\hline
0.004 & 0.084/0.085 & 1.1 & 92 \cite{ReEA1992} \\
\hline
0.202 & 0.37/0.41 & 1.5 & 96 \cite{UeEA1996} \\
\hline
0.202 & 0.37/0.41 & 1.6 & 98 \cite{UeEA1998} \\
\hline
20 & 2.2/3.8 & 30 & 99 \cite{SaEA1999} \\
\hline
\end{tabular}
\caption{Summary of cavity loss measurements. Lpm = Loss per mirror.}
\end{center}
\label{tab:cavityloss}
\end{table}
Table \ref{tab:cavityloss} summarizes previous loss measurements on cavities of various sizes. Round-trip loss in the km-scale arm cavities of the second-generation GW detectors not including the input transmission is expected to be of order 100\,ppm. These values indicate that short cavities typically have lower loss than long cavities. One of the questions is why this is the case.

The main loss mechanism in these cavities is scattering from the optics. The quality of currently available optics does not seem to be sufficient to allow us to build weak scattering (low-loss) filter cavities. Scattering is caused by point defects on the mirror surface or inside the mirror coating \cite{BeEA2004,Hir2010}, by residual surface roughness after substrate polishing and coating \cite{WLO1999,Bla2002,ZaEA2008}, and by larger-scale figure errors \cite{GTF1990}. Ultimately, the loss contribution of each scattering process needs to be known and linked to some deficiency of the coating and substrate fabrication process. 

Figure errors and surface roughness are currently the dominant cause of scattering in long cavities. The reason is that the beam size on the optics increases with cavity size and therefore the beam is sensitive to larger-scale mirror errors. Errors at larger scales are typically worse than small-scale errors. Decreasing the beam size, losses get smaller until one eventually comes into the regime where scattering is dominated by point defects. The density of point defects is often very high and a beam is scattered from many defects simultaneously. In contrast, the very small loss values in table \ref{tab:cavityloss} can only be realized by steering a very small beam into a region of the mirror surface that has no significant defects. The beam size in a 50\,m cavity would be about 8\,mm, which is larger than the typical distance between defects on current optics. If this is truly the case, then the only solution to the scattering problem in filter cavities is to improve current coating and substrate technology. As a first step to understand quality limits in the fabrication of optics, we measured scattering from two high-quality optics as a function of scattering angle, which allows us to draw further conclusions about the scattering process involved. The optics made of superpolished fused silica substrates with silica:tantala dielectric coatings represent the current state-of-the art optical technology that could be used in filter cavities.

\section{Experiment} 

\subsection{Samples}

\begin{figure}[htb]
\centerline{\includegraphics[width=8.7cm]{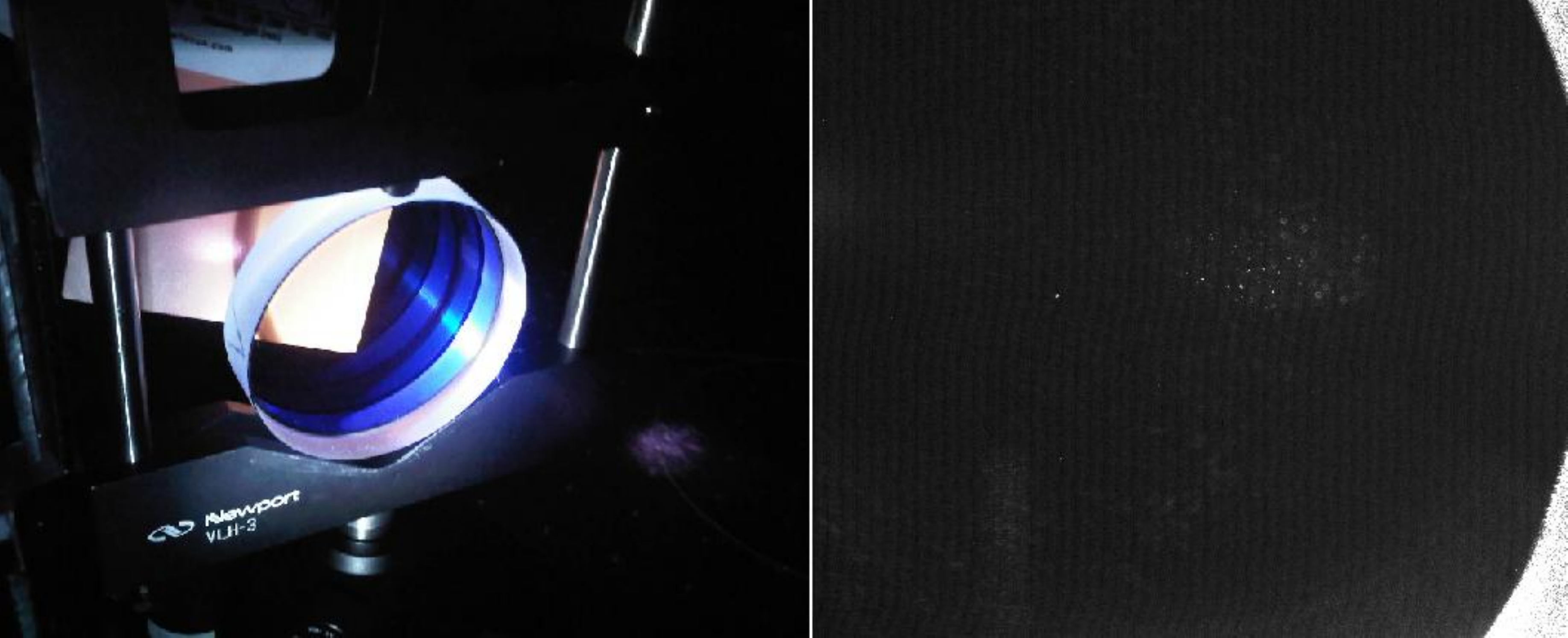}}
\caption{The beamsplitter shown in room light (left) and as viewed by our setup with 1064\,nm laser illumination (right).}\label{fig:bs}
\end{figure}

\begin{figure}[htb]
\centerline{\includegraphics[width=8.7cm]{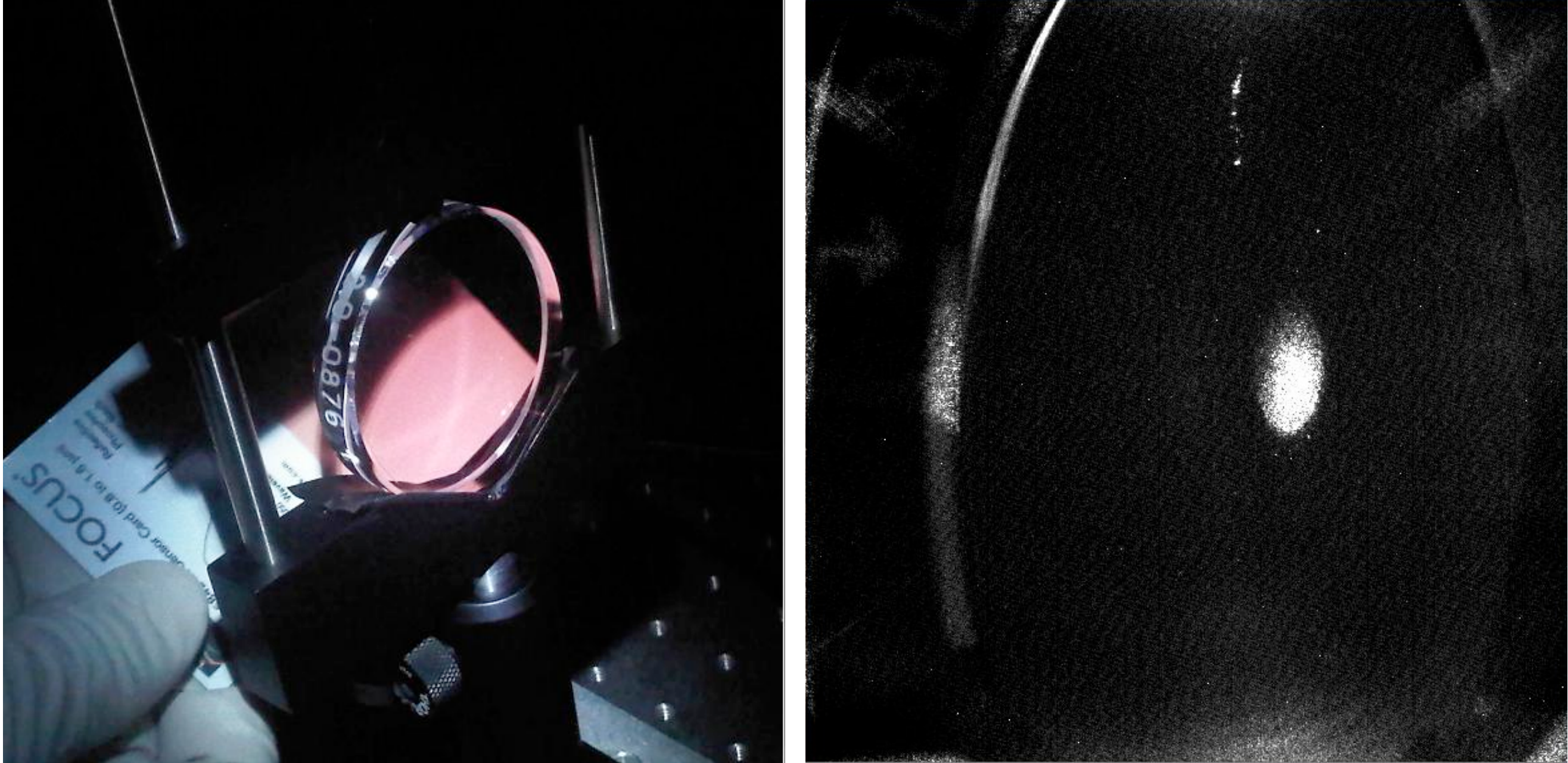}}
\caption{The highly-reflective mirror shown in room light (left) and as viewed by our setup with 1064\,nm laser illumination (right).}\label{fig:hrm}
\end{figure}

The optics we measured were both two-inch diameter fused silica substrates, with superpolished surfaces, and ion-sputtered dielectric coatings (see, e.g. chapter 2 in~\cite{har2012}) composed of alternating layers of silica and tantala. 

The beam splitter (BS) we measured is shown in Figure~\ref{fig:bs}. It was coated by Advanced Thin Films of Boulder CO for use at 1064\,nm and 532\,nm wavelengths and with a 45 degree nominal angle of incidence. We measured an 18.4\% reflectivity and 81.4\% transmission for horizontally-polarized 1064\,nm light incident at 45 degrees. The barrel of this optic is unpolished and it scatters brightly even for a small beam incident on the center of the optical surface. The bright scatter from the barrel is also mirrored by the front and back surfaces of the optic. 

The highly-reflecting mirror (HRM) is shown in Figure~\ref{fig:hrm}. It was manufactured by Gooch and Housego (S/N 13589) for operation at 1064\,nm wavelength. The transmission of the HRM was measured using a 1064\,nm laser and calibrated power meter to be roughly 20 ppm from 0-35 degrees for p-polarization. The barrel of this optic is polished, and scattered much less than that of the BS. 

\subsection{Experimental setup}

\begin{figure}[htb]
\centerline{\includegraphics[width=8cm]{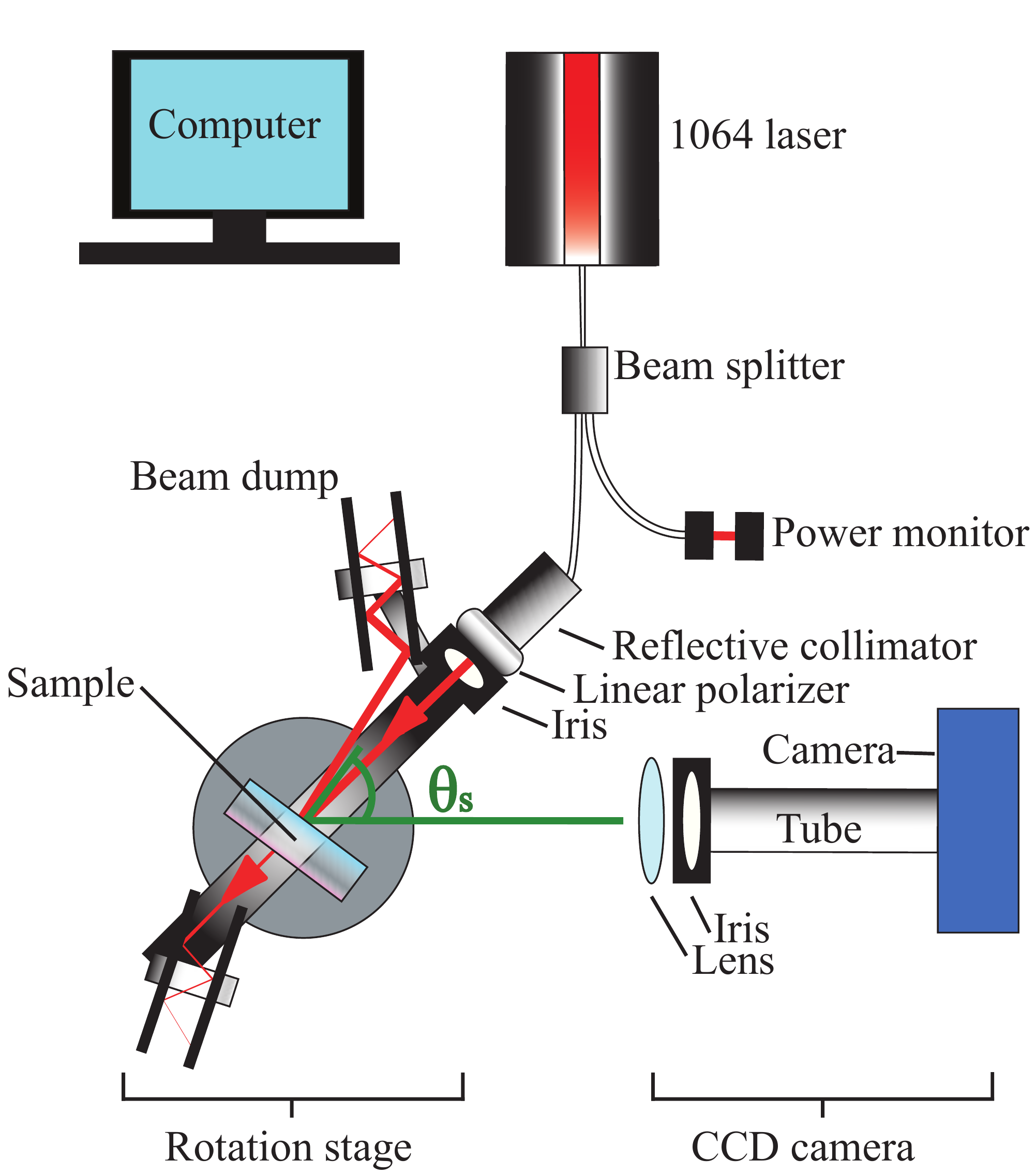}}
\caption{The layout of the imaging scatterometer.}\label{layout}
\end{figure}

Figure~\ref{layout} shows the layout of the imaging scatterometer that was used to characterize the forward-scattering of laser light from our optics. Its basic operation is as follows. A laser beam illuminates the optical sample at a fixed incidence angle. Images of the light scattered from the sample's surface(s) are recorded over a discrete set of scattering angles, $\theta_s$, defined as the angle between the camera's imaging direction and the normal to the sample surface~\cite{Stover1995}. The scattering angle is adjusted using a motorized rotation stage. The angles, with respect to the incident beam, that are accessible by this setup range from the smallest angle that allows the imaging optics to see past the fiber launch (5 degrees) to the angle where the optical surface is parallel to the CCD camera (90 degrees for normal incidence). A consequence of this is that when the angle of incidence on the sample is near normal, only large angle scattering can be measured.  

The incident laser power is provided by a linearly polarized, 1064\,nm nominally 500\,mW laser (model: CrystaLaser CL 1064-500-SO). The laser beam is guided to the setup by an optical fiber that includes a 90:10 beamsplitter. The 10\% fiber is directed to a photodiode for power monitoring, while the 90\% fiber is directed to a series of optics mounted on the rotation stage. When the laser exits the narrow fiber aperture it is strongly diverging, so we first collimate it to an 8.5\,mm beam diameter using a reflective collimator (Thorlabs RC08FC). Then a linear polarizer is used to further improve the laser polarization and ensure that it is horizontal (parallel to the tabletop). Finally, an adjustable iris is set to approximately 6\,mm diameter to clip the beam edges and reduce the amount of light falling on the sample holder, which has a much higher scatter coefficient than the samples, and can contaminate the images. The beam incident on the sample is thus horizontally polarized with approximately 8\,mm diameter waist, truncated to a diameter of 6\,mm, and has about 150\,mW of power. 

The optical sample, the fiber-coupled laser output, and two black-glass beam traps are mounted on a motorized rotation stage. This ensures that as the stage is rotated the angle of incidence remains fixed and the reflected and transmitted beams are dumped. The center of the front surface of the optic is positioned directly above the rotation axis of the stage so that when the stage is rotated the optic does not translate in the images.

The imaging system views the optical surface at an angle with respect to the input beam axis. A single two-inch diameter f=100\,mm bi-convex lens and adjustable iris with diameter set to 16\,mm are located about 50\,cm from the sample and form an image of the optical surface on a 1 mega-pixel astronomical Charged-coupled device (CCD) camera (Apogee Alta U6). Between the iris and the camera an aluminum tube with an RG850 optical high pass filter at its entrance is used to reduce room lights and unwanted stray laser light from reaching the camera. 

Measurements are performed by taking a camera exposure at a given scattering angle, while also recording the input laser power, then rotating the stage a small amount (such as 1 degree) and repeating. The camera exposure times are adjusted to provide a good scattered light signal-to-noise ratio, while not saturating parts of the image close to the laser beam spot on the sample surface(s). For angles near to the intense specular reflection from the BS, the camera saturates even at its shortest exposure time (0.02 s). These angles have been removed from our analysis. 

Prior to measurement, the front and back surfaces of each sample were drag-wiped with an optical tissue and methanol to remove dust and impurities. The incidence angle was set to 45 degrees for the BS and 3 degrees, the smallest angle that allowed dumping the back-reflected beam cleanly in the black-glass trap, for the HRM. The setup is housed in a soft-wall clean room that uses overpressure and laminar flow to reduce airborne particles. 

\subsection{Calibration}

\begin{figure}[htb]
\centerline{\includegraphics[width=8.7cm]{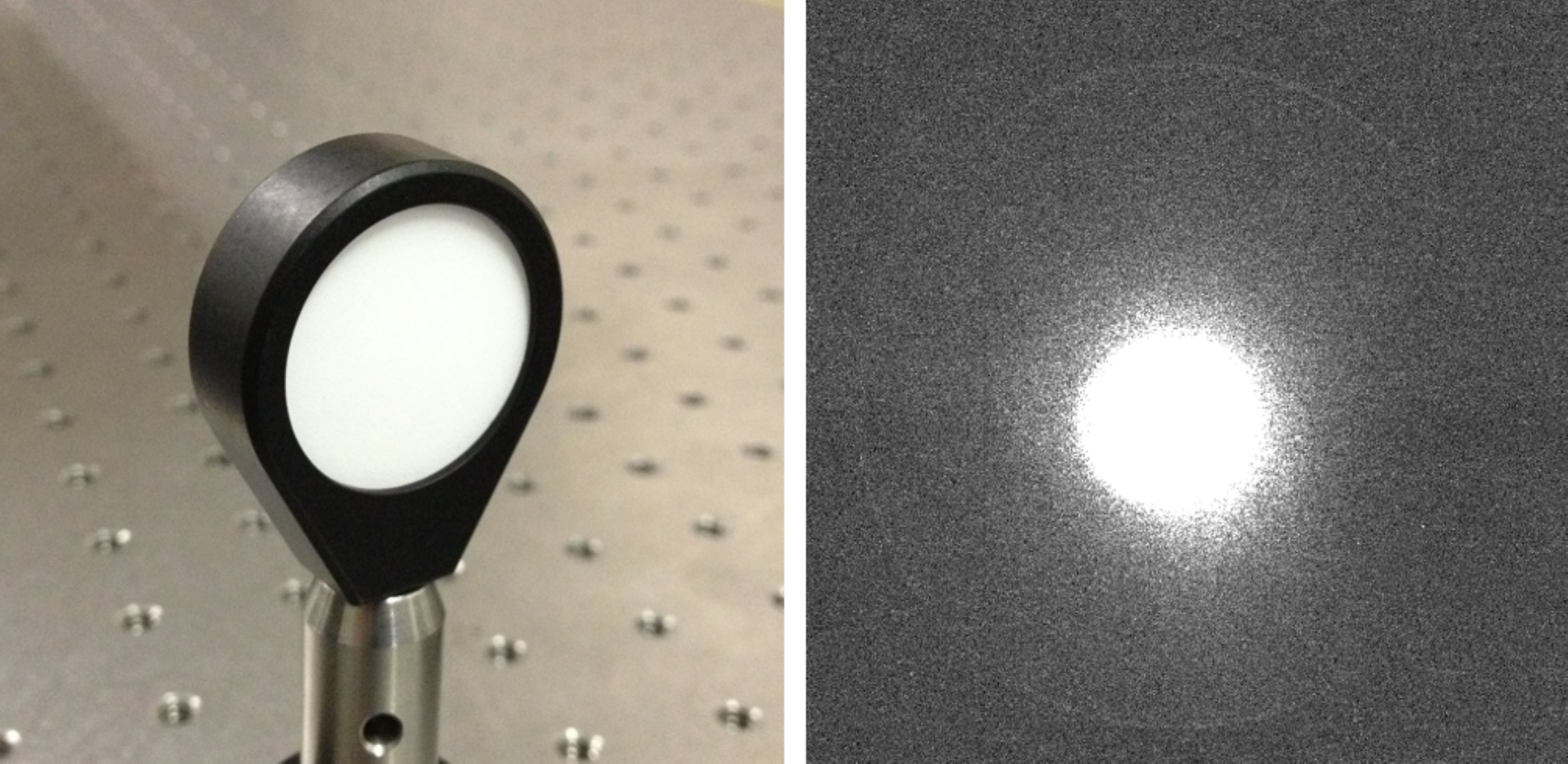}}
\caption{The diffusing sample shown in room light (left) and as viewed by our setup with 1064nm laser illumination (right).}\label{fig:diffuser}
\end{figure}

For our analysis we need to convert the image counts measured by the CCD camera into a calibrated measure of scatter. To do this we measure the scattered light from a diffusing sample twice, once with the CCD camera and once with a calibrated power meter. We then compare their readings. This uses the setup as described above, but with i) a diffusing target (Spectralon Diffusion Material, 1" x 0.012" disk, SM-00875-200) as the sample \cite{Bruegge2001, BhEA2011}, ii) an additional power meter that can be inserted in front of the imaging optics to measure scattered light power, and iii) a neutral-density filter with a measured 1/273 transmission factor at 1064\,nm inserted in front of the CCD camera to prevent saturation. Images of the diffusing sample in room light and with incident laser light and viewed via the CCD camera are shown in Figure~\ref{fig:diffuser}. 

For these calibrations, the power meter was located 21.6\,cm from the center of the sample surface, and at the height of the laser beam (which is parallel to the tabletop). With it we recorded the scattered power over a set of scattering angles separated by 10 degree steps. From these measurements we calculated the standard bi-directional reflectance distribution function~\cite{Stover1995},
\begin{equation}\label{brdf}
BRDF=\frac{P_s}{P_i \Omega \cos{\theta_s}}
\end{equation} 
where $P_i$ is the incident laser power, $P_s$ is the scattered light power detected by the power meter, which subtends a solid angle $\Omega$, and is oriented at polar angle $\theta_s$ with respect to the normal to the optical surface (in the plane of the laser beam). 

Then we used the CCD camera to take images for the same scattering angles. For each image, a region of interest (ROI) corresponding to the area of the sample surface with significant light power incident on it is chosen. The counts over the entire ROI are summed and normalized by the camera exposure time $T_{exp}$ and incident laser power on the sample to give, 
\begin{equation}\label{arbccd}
ARB_{CCD}=\frac{\sum_k V_k}{T_{exp} P_{i}}
\end{equation}
where $V_k$ is the value of the $k$th pixel in the ROI. 

We use the fact that the BRDF is intrinsic to the sample to calculate a calibration function, 
\begin{equation}\label{fcal}
F=\frac{BRDF \cos(\theta_s)}{ARB_{CCD}},
\end{equation}
that relates the CCD counts to the BRDF measured by the power meter. 

\begin{figure}[htb]
\centerline{\includegraphics[width=8.7cm]{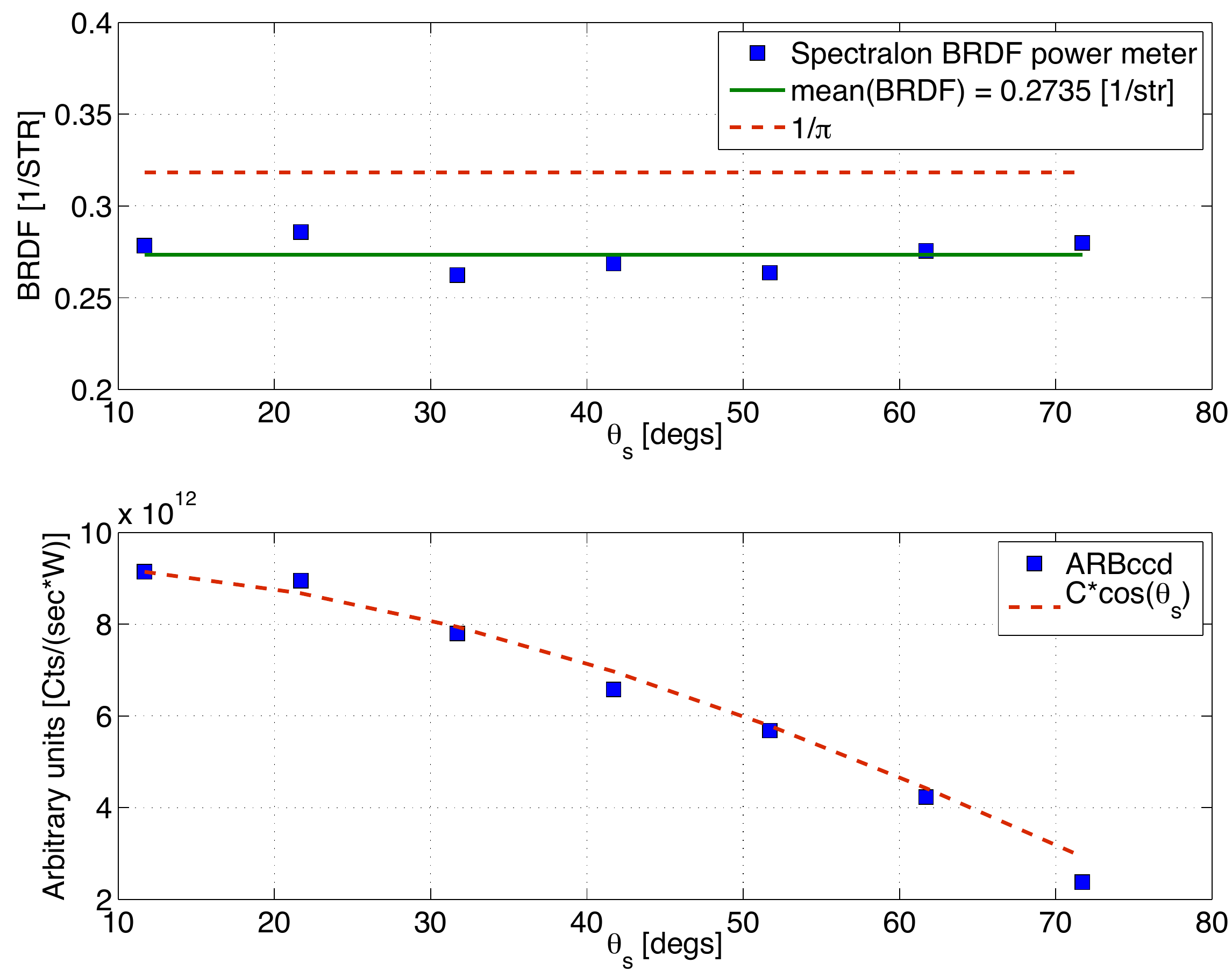}}
\caption{The top graph shows BRDF measured by the power meter and a comparison with the $1/\pi$ value expected for an ideal Lambertian diffuser (uniform hemispherical scattering, see Chapter 11 in~\cite{har2012}). The bottom graph shows the counts recorded by the CCD and normalized by the exposure time and incident power. This is compared to the $\cos(\theta_s)$ scattered power dependency expected for an ideal Lambertian diffuser.}\label{fig:cal-brdf}
\end{figure}

Figure~\ref{fig:cal-brdf} shows the results of the calibration used for this paper. We measured a mean BRDF of 0.27 STR$^{-1}$ for our diffuse scattering sample for near normal incidence. This value is consistent with BRDF measured for similar targets~\cite{Bruegge2001, Cohen2003}. The normalized counts measured by the CCD behave as we would expect, falling off as the cosine of the scattering angle. Following equation~\ref{fcal} we calculate a calibration constant of $F=3.20\times10^{-14}$ W\,sec\,Counts$^{-1}$\,STR$^{-1}$. This value is used in the following analysis. 

\subsection{Data analysis}

\begin{figure}[htb]
\centerline{\includegraphics[width=8cm]{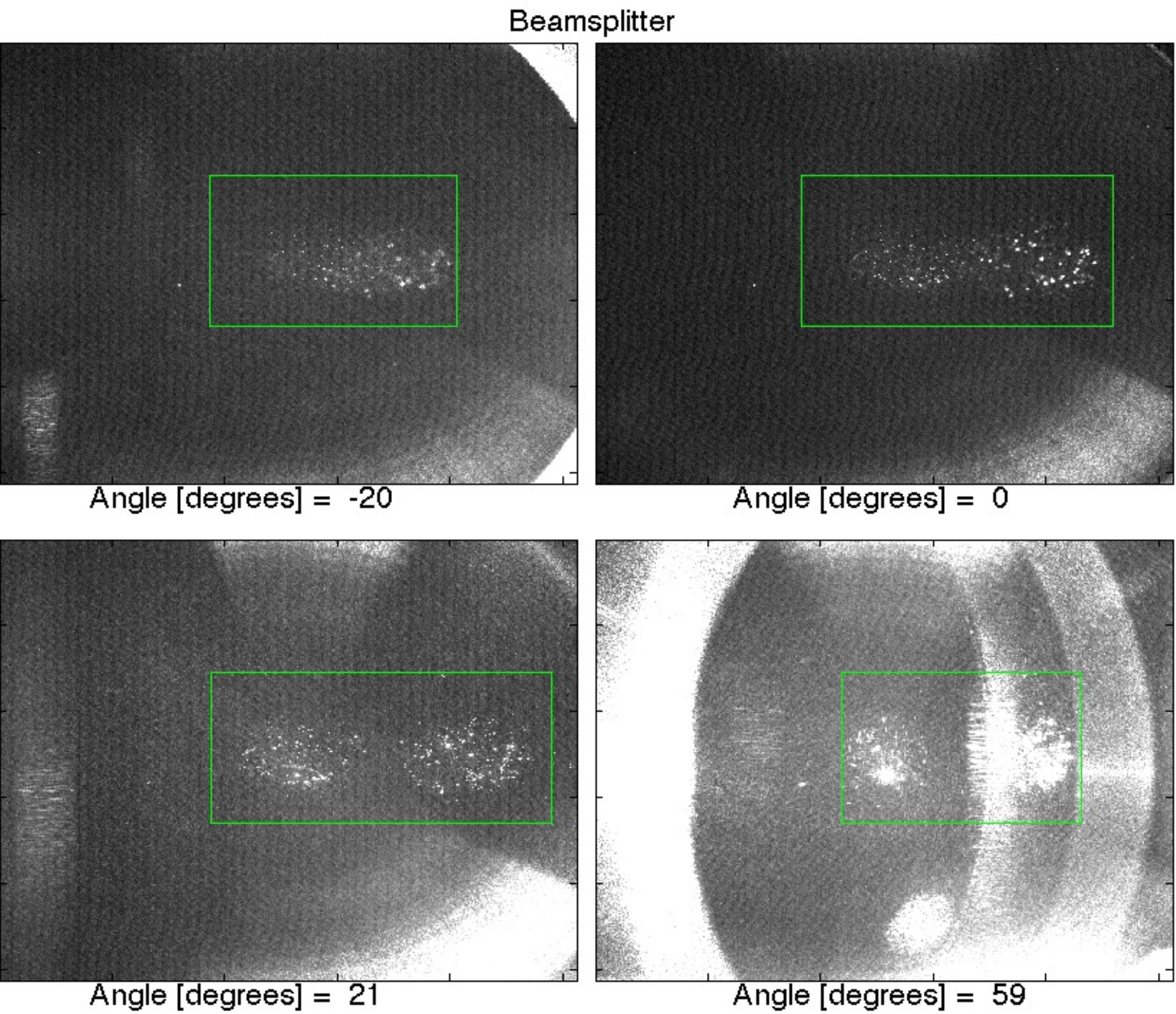}}
\caption{The beam splitter viewed at scattering angles of -20 degrees (upper left), 0 degrees (upper right), 21 degrees (lower left), and 59 degrees (lower right). The rectangles are the regions of interest that are used to measure the scattered light.}\label{fig:bs_roi}
\end{figure}

\begin{figure}[htb]
\centerline{\includegraphics[width=8cm]{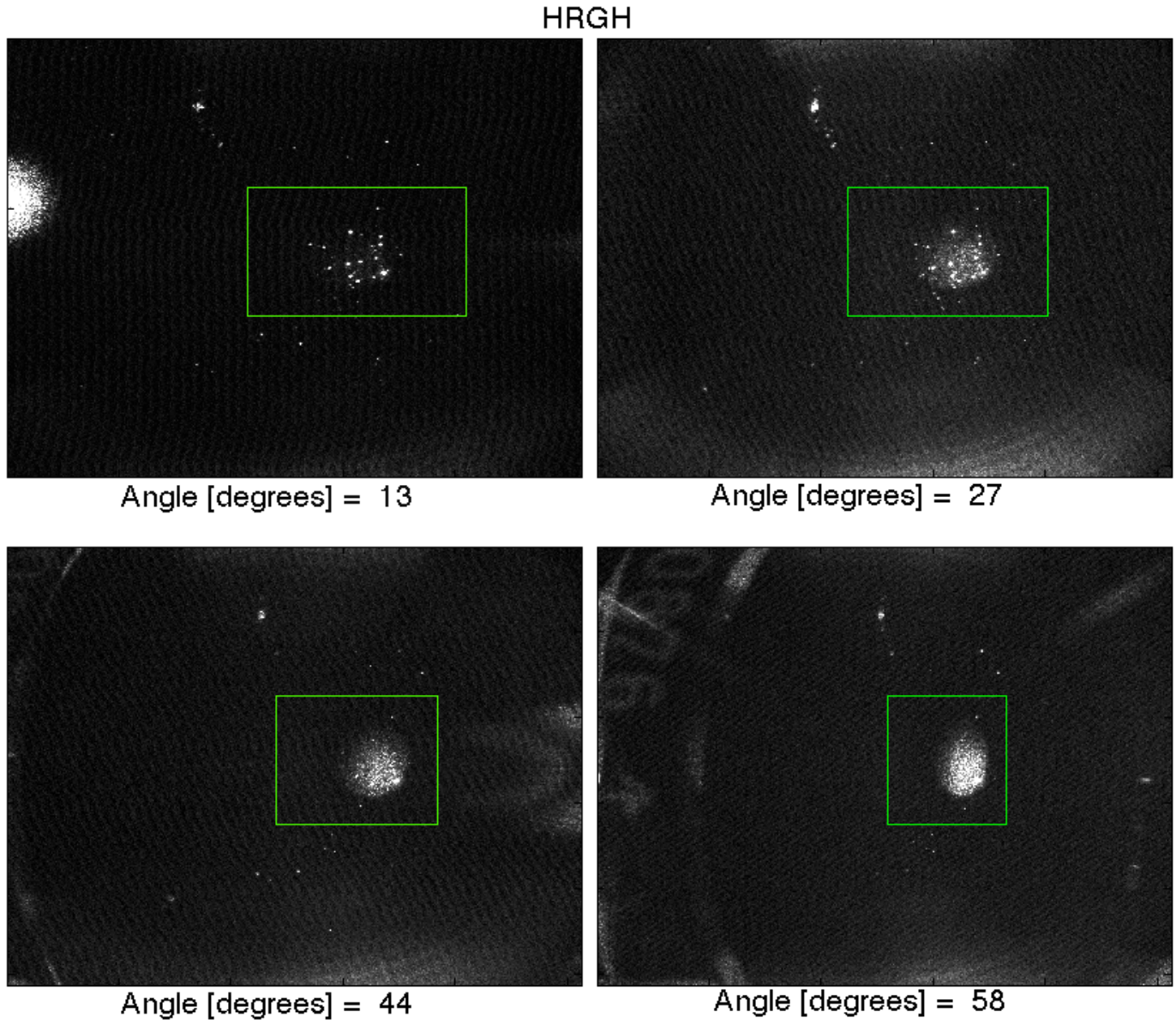}}
\caption{The highly-reflective mirror viewed at scattering angles of 13 degrees (upper left), 27 degrees (upper right), 44 degrees (lower left), and 58 degrees (lower right). The rectangles are the regions of interest that are used to measure the scattered light.}\label{fig:hrm_roi}
\end{figure}

As described above, the data products measured for each sample are megapixel CCD images taken over a range of discrete scattering angles. The procedure for analyzing the images and producing BRDF curves is as follows. First a dark image (an image with the same exposure time, but with the laser turned off) is subtracted from each image, removing the camera noise and any hot pixels. Then we select for each image a region of interest, a rectangular area that captures the vast majority of the light scattered by the surface of the optic, while including as little as possible of the scattering from the optic barrel or the optic mount. A region of interest encompassing scatter from both front and back surfaces was used for the BS, because significant amounts of scattered light was visible from each surface (the fact that the scatter from the back surface is diminished by two passes through the 81.4\% reflectivity front surface was not taken into account). The scatter from the HRM was dominated by scatter from the front surface so only that was included in the region of interest. Then we sum the pixel values in the region of interest of the subtracted image to calculate $ARB_{CCD}$, following Equation~\ref{arbccd}, and finally calibrate the values into BRDF using the calibration constant given above, and the relation, 
\begin{equation}\label{brdfccd}
BRDF=ARB_{CCD} F.
\end{equation}

Figure~\ref{fig:bs_roi} shows a set of four example subtracted images for the BS. The scatter from the front and back surfaces can be seen to separate as the sample is rotated through scattering angles from -20 to 59 degrees. The region of interest is indicated with a rectangular line. Note that the unpolished barrel of the BS scatters brightly, and that it's reflections off the mirror surfaces pollute the region of interest for large viewing angles. Figure~\ref{fig:hrm_roi} shows four separate scattering images over a range of angles from the HRM. Here there is much less background scatter with respect to the BS because the polished barrel of the HRM scatters far less. Also, the regions of interest encompass only the front surfaces. As the scattering angle gets larger the scattering character changes from a constellation of point scatterers to a diffuse glow.  

Figures~\ref{fig:bs_roi} and \ref{fig:hrm_roi} show two distinct types of scattering. Where the beam is most intense on the sample surface, a strong central scattering 'glow' is seen. This has a diffuse speckle pattern that is characteristic of scatter caused by surface roughness. In this region, and even well outside of this region where the laser power is significantly less, a sparse constellation of bright points can be seen. We find that some of these points can be removed or relocated by drag-wiping, indicating that they originate from dust and other impurities on the surface. The images shown correspond to the cleanest data set over several drag wipes, performed in situ in our clean room. 

\section{Results} 

\begin{figure}[htb]
\centerline{\includegraphics[width=8.7cm]{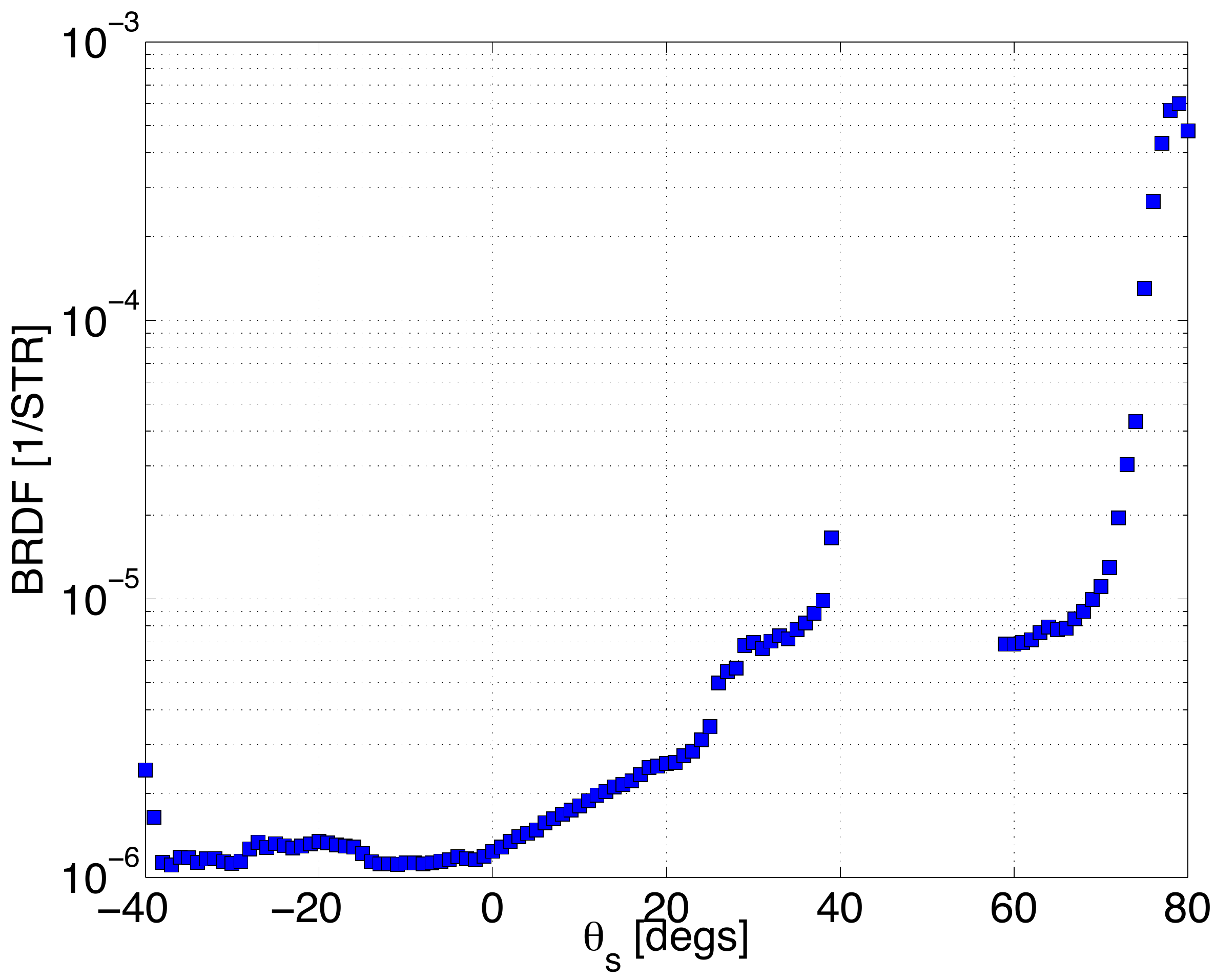}}
\caption{BRDF versus scattering angle for the beamsplitter sample. Data points have been removed for angles where the CCD camera saturated due to the strong near-specular reflection at 45 degrees. The BRDF from 60 to 80 degrees is increasingly overestimated due to spatial overlap with scatter from the unpolished barrel, see lower right image in Figure~\ref{fig:bs_roi}.}\label{fig:brdf_bs}
\end{figure}

Figure~\ref{fig:brdf_bs} shows the BRDF results for the beam splitter (45 degrees angle of incidence) versus scattering angle. This data can be separated into a few distinct regions. In the -40 to 40 degree region, the images were similar to the first three panels shown in Figure~\ref{fig:bs_roi}. Scatter from the front and back surfaces were imaged cleanly without influence from unwanted stray light. Here we see BRDF with values between $1\times10^{-6}$ and $1\times10^{-5}$ STR$^{-1}$. In the range from 40 to 58 degrees the CCD camera was saturated by light from the beamsplitter's specular reflection. This data has been removed. The data from 60 to 80 degrees includes scatter from the BS surfaces, but this is polluted by an increasingly large component of unwanted scattered light from the unpolished barrel of the optic and the optical mount. This BRDF in this region is not to be trusted. 

\begin{figure}[htb]
\centerline{\includegraphics[width=8.7cm]{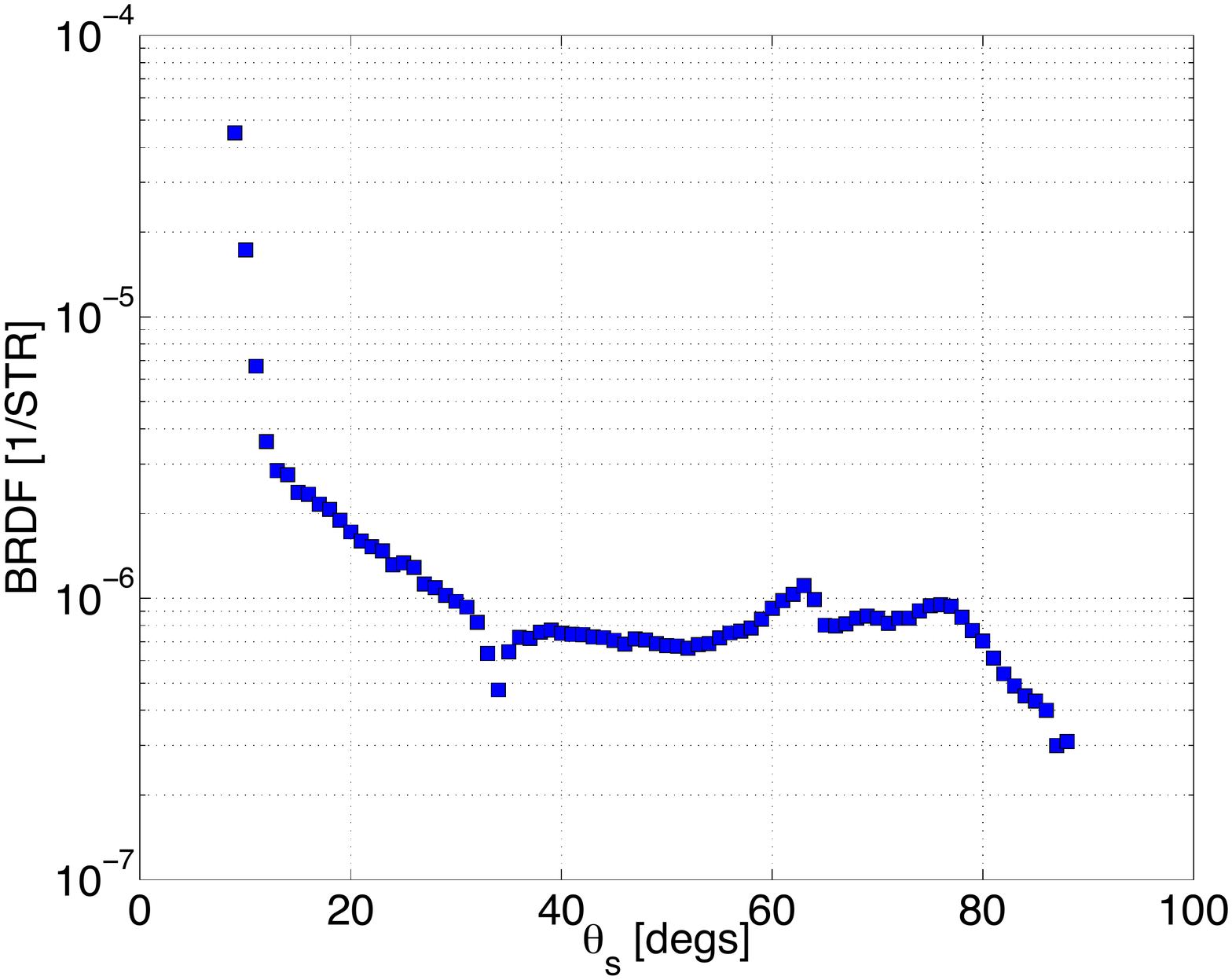}}
\caption{BRDF versus scattering angle for the highly-reflecting mirror.}\label{fig:brdf_hrm}
\end{figure}

Figure~\ref{fig:brdf_hrm} shows the measured BRDF for the highly-reflecting mirror. This data was significantly cleaner than for the BS due to less unwanted scattered light polluting the images. The BRDF for this mirror is quite low, below 10$^{-6}$ for scattering angles greater than 30 degrees. The BRDF for the HRM was checked using a similar scatterometer, which utilized a single photodetector and chopper wheel, in another lab. For normal incidence we measured BRDF$=6\times10^{-7}$ STR$^{-1}$ at $\theta_s=45$ degrees, which is close to the value for that angle in Figure 9.

These BRDF values can also be used to estimate the total scattering associated with these optics. Total integrated scatter (TIS)~\cite{Stover1995} can be estimated by integrating a measurement of BRDF times $\cos\theta_s$ over the full solid angle of scatter (a hemisphere for back-scatter), to get the hemispherical reflectance~\cite{bely03}, 
\begin{equation}\label{eqn:hemref}
R_H = \frac{P_s}{P_i} = R\,\rm{TIS} = \int_0^{2\pi}\int_0^{\pi/2} \rm{BRDF} \cos{\theta_s}\sin{\theta_s}d\theta_s d\phi_s,
\end{equation}
where\footnote{For a perfect Lambertian diffuser, BRDF is constant and $R$ and TIS are unity. Evaluating equation~\ref{eqn:hemref} with this information leads to, BRDF$=1/\pi$ for all forward scattering angles, the value referred to above~\cite{har2012}.} $R$ is the reflectivity of the optic, and $\phi_s$ are the azimuthal scattering angles. 

We assume independence of the BRDF function on azimuthal scattering angles, since we use smoothly polished optics that should scatter isotropically at a given polar angle. This allows integration of the scatterometer data over only the polar angle of scatter. We approximate the solid angle integral as a sum of the scatter in individual rings centered on the polar angles of measured scatter $\theta_s$,
\begin{eqnarray}
R\,\rm{TIS}(\theta_s) &=& \Omega_{\rm{ring}}(\theta_1, \theta_2) \rm{BRDF}(\theta_s)\cos\theta_s \\
&=& 2\pi(\cos\theta_1-\cos\theta_2) \rm{BRDF}(\theta_s)\cos\theta_s,
\end{eqnarray}
where $\Omega_{\rm{ring}}=2\pi(\cos{\theta_2}-\cos{\theta_1})$ is the solid angle of the ring subtended by polar angles between $\theta_1 = \theta_s - d\theta/2$ and $\theta_2 = \theta_s + d\theta/2$, where $d\theta$ is the angular resolution (step) of the scatterometer measurement.

Table~\ref{tab:tis} shows the total integrated scatter values calculated for the BS and HRM using the data presented in Figures~\ref{fig:brdf_bs} and \ref{fig:brdf_hrm} and the equations above. 

\begin{table}[ht!]
\begin{center}
\begin{tabular}{|l|l|l|}
\hline
Sample & Angle range [degrees]  & R TIS [ppm] \\
\hline\hline
BS & [-40,0] & 1.62 \\
\hline
BS & [0,37] & 5.71 \\
\hline
BS & [60,80] & 57.0 \\
\hline
HRM & [9,85] & 3.79 \\
\hline
\end{tabular}
\caption{Total integrated scatter calculated from BRDF measurements in Figures~\ref{fig:brdf_bs} and \ref{fig:brdf_hrm}. The 60-80 degree range for the BS is polluted by scatter from the optic barrel.}
\end{center}
\label{tab:tis}
\end{table}

\section{Discussion} 
The BRDF measurement of a high-reflective optic showed that the total integrated light scattered into larger angles can be as small as 4\,ppm provided that the optic is cleaned and the measurement is carried out in a near particle-free environment. Somewhat larger BRDF values were measured for a beamsplitter, but as discussed in the text, it is also more challenging to obtain unbiased results for BRDF measurements on optics with non-negligible transmission and unpolished surfaces. 

Large-angle scattering of superpolished optics is generally associated with point defects in mirror coatings. The TIS values shown in Table~\ref{tab:tis} need to be added to estimates of small-angle surface-roughness scattering that is usually obtained from measured surface-roughness profiles. These profiles are unknown for the two optics measured for this paper, but extrapolating estimates from measured surface-roughness spectra as for example presented in \cite{Kel2009}, one obtains additional scattering of 1\,ppm and less. It should be noted though that the same article points out that a discrepancy has been observed in the past between measured BRDF and estimates obtained from surface-roughness spectra. Nevertheless, based on the values at hand, it seems possible that filter cavities with round-trip scatter loss less than 20\,ppm can be constructed with the available optics. 

The main goal in the near future will be to form a consistent picture between BRDF measurements, scattering estimates from surface-roughness measurements, and actual loss observed in optical cavities for example through ring-down measurements. Further measurements and simulations are necessary to explore the small-angle regime (below 10$^\circ$) and improve our scatter-loss predictions. Even though filter cavities will have lengths of 50\,m and longer, loss measurements on smaller cavities can help to improve our understanding of scattering provided that cavity-loss measurements are combined with scattering measurements on individual optics that form the cavity. 

\section*{Acknowledgments}
This work is supported by the Research Corporation for Science Advancement Cottrell College Science Award \#19839 and by National Science Foundation Awards PHY-0970147, PHY-0854812, and PHY-0555406. We thank our colleagues in the LIGO Scientific Collaboration for fruitful discussions about this research and for review of this manuscript. 

\end{document}